\documentclass{PoS}
\usepackage{epsfig}
\usepackage{bm}
\usepackage{amssymb}
\usepackage{fontenc}
\usepackage{times}
\usepackage{mathptmx}
\usepackage{graphicx}

\title{Searching for the elusive critical endpoint at finite temperature and 
isospin density}

\ShortTitle{Searching for the elusive critical endpoint at finite temperature 
and isospin density}

\author{\speaker{D.~K.~Sinclair}\\
HEP Division, Argonne National Laboratory, 9700 South Cass Avenue, Argonne,
IL 60439, USA\\
        E-mail: \email{dks@hep.anl.gov}}

\author{J.~B.~Kogut\\
Department of Energy, Division of High Energy Physics, Washington, DC 20585,
USA\\
and\\
Dept. of Physics -- TQHN, Univ. of Maryland, 82 Regents Dr., College
Park, MD 20742, USA\\
        E-mail: \email{jbkogut@umd.edu}}

\abstract{We consider 3-flavour lattice QCD with a finite chemical potential
$\mu_I$ for isospin, close to the finite temperature transition from hadronic
matter to a quark-gluon plasma. In this region one can argue that the position
and probably the nature of this transition mimic those at finite quark-number
chemical potential $\mu$. The quark mass is chosen to be close to the critical
mass at zero chemical potentials. Since the Binder cumulants used to determine
the nature of this transition in HMD(R) simulations are very sensitive to the
updating increment $dt$, we have switched to the newer exact RHMC algorithm
for our simulations. Preliminary results indicate that there is no critical
endpoint in the small $\mu_I$ regime, at least none connected with the critical
point at zero chemical potentials.}

\FullConference{XXIVth International Symposium on Lattice Field Theory\\
                July 23-28, 2006\\
                Tucson, Arizona, USA}
                                                                                
\begin{document}

\section{Introduction}

Direct simulations of QCD at finite baryon/quark number density are made
difficult if not impossible because at finite quark-number chemical potential
$\mu$ the fermion determinant is complex. At small chemical potentials, close
to the finite temperature transition, various methods have been devised to
circumvent this difficulty, series expansions in $\mu$
\cite{Allton:2002zi,Allton:2005gk,Gavai:2003mf}, analytic continuation from
imaginary $\mu$ \cite{deForcrand:2002ci,deForcrand:2006pv,Azcoiti:2004ri,%
Azcoiti:2005tv, D'Elia:2002gd,D'Elia:2004at}, reweighting methods
\cite{Fodor:2001au,Fodor:2004nz} and canonical ensemble techniques
\cite{Kratochvila:2005mk,Alexandru:2005ix}.
 
We adopt a different strategy, and simulate using the magnitude of the fermion
determinant and ignoring the phase \cite{Kogut:2002zg,Kogut:2005yu}. This can
be thought of as considering all quarks to be in isodoublets and introducing a
finite chemical potential $\mu_I$ for isospin. In the region of small
$\mu/\mu_I$, where the phase is expected to be less important one can argue
that the finite $\mu$ and $\mu_I$ transitions might be identical. Since our
fermion determinant is positive (or at least non-negative), we can use
standard hybrid molecular-dynamics HMD(R) simulations \cite{Gottlieb:1987mq}.
However, for this algorithm, the Binder cumulants used to determine the nature
of the finite temperature transition turn out to be strongly dependent on the
updating increment $dt$. For this reason we now simulate using the rational
hybrid monte-carlo (RHMC) algorithm \cite{Kennedy:1998cu,Clark:2005sq}, which
is exact in the sense of having no $dt$ dependence for observables.

In the low chemical potential domain, the most interesting feature expected in
the phase diagram is the critical endpoint, where the finite temperature
transition changes from a crossover to a first-order transition as chemical
potential is increased. The critical endpoint is expected to lie in the
universality class of the 3-dimensional Ising model. For 3 flavours it had
been expected that the critical point at zero chemical potentials, where the
transition changes from a first order transition to a crossover as mass is
increased, would move to higher masses as the chemical potential increases,
thereby becoming the critical endpoint. Our preliminary results indicate that
this does not happen. 
  
In section~2 we give the fermion action and make a few comments on the RHMC
implementation. Section~3 gives our preliminary results. Our conclusions
occupy section~4.

\section{QCD at finite isospin density and the RHMC}

The pseudo-fermion action for QCD at finite $\mu_I$, used for the implementation
of the RHMC algorithm is
\begin{equation}
S_{pf}=p_\psi^\dag {\cal M}^{-N_f/8} p_\psi
\label{eqn:action}
\end{equation}
where $p_\psi$ is the momentum conjugate to the pseudo-fermion field $\psi$.
\begin{equation}
{\cal M} = [D\!\!\!\!/(\frac{1}{2}\mu_I)+m]^\dag
           [D\!\!\!\!/(\frac{1}{2}\mu_I)+m] + \lambda^2
\end{equation}
is the quadratic Dirac operator, and we set $\lambda=0$ for our $\mu_I < m_\pi$
simulations.

To implement the RHMC method we need to know positive upper and lower bounds
to the spectrum of ${\cal M}$. $25$ exceeds the upper bound for the $\mu_I$
range of interest. We use a speculative lower bound of $10^{-4}$ since the
actual lower bound of the spectrum is unknown. This is justified by varying
the choice of lower bounds and comparing the results \cite{Kogut:2006jg}. For
$N_f=3$ we use a $(20,20)$ rational approximation to ${\cal M}^{(\pm 3/16)}$
at the ends of each trajectory, and a  $(10,10)$ rational approximation to
${\cal M}^{(-3/8)}$ for the updating.

\section{Simulations and Results}

We are simulating lattice QCD with staggered fermions and $N_f=3$ at quark
masses close to $m_c$, the critical mass for $\mu=\mu_I=0$ on $8^3 \times 4$,
$12^3 \times 4$ and $16^3 \times 4$ lattices. $m=0.02$, $0.025$, $0.03$,
$0.035$, and $\mu_I=0.0,\,0.2,\,0.3$. For our $12^3 \times 4$ simulations we
use runs of 300,000 trajectories at each of 4 $\beta$ values close to
$\beta_c$, for each $m$ and $\mu_I$. We mostly use $dt=0.05$ for which
length-1 trajectories give acceptances of $\sim 70\%$ for the RHMC algorithm.

To determine the nature of the transition, we use 4-th order Binder cumulants
\cite{Binder:1981sa} for the chiral condensate. For any observable $X$ this
cumulant is defined by
\begin{equation}
B_4(X) = {\langle(X-\langle X \rangle)^4\rangle \over
          \langle(X-\langle X \rangle)^2\rangle^2}
\end{equation}
where the $X$s are lattice averaged quantities. For infinite volumes, $B_4=3$
for a crossover, $B_4=1$ for a first-order transition and $B_4=1.604(1)$ for
the 3-dimensional Ising model. Thus, if there is a critical endpoint we would
expect $B_4$ to decrease with increasing $\mu_I$, passing through a value close
to the Ising value at the critical $\mu_I$. 

\begin{figure}[htb]
\epsfxsize=4in
\centerline{\epsffile{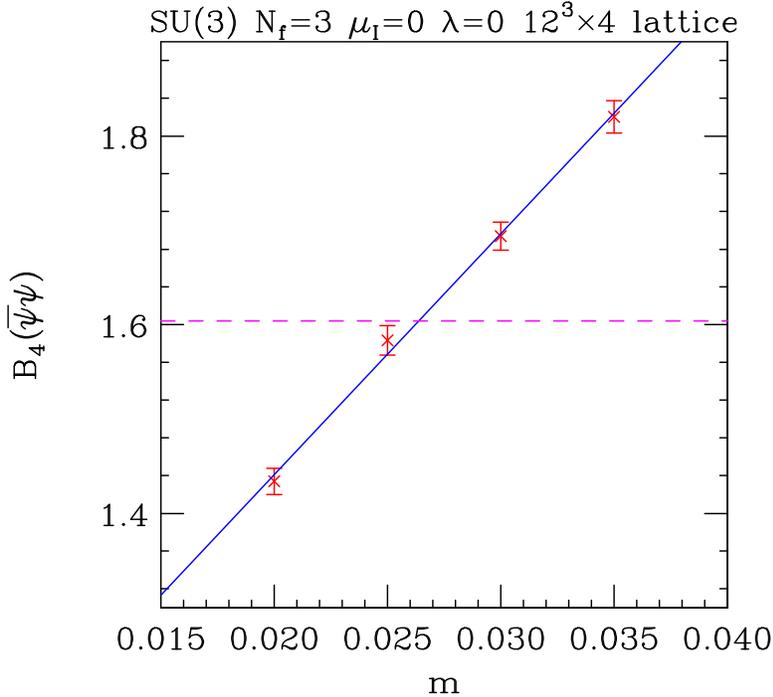}}
\caption{Binder cumulant at $T=T_c$ as a function of mass at $\mu_I=\mu=0$.}
\label{fig:b4mass}
\end{figure}
Figure~\ref{fig:b4mass} shows our preliminary measurements of the Binder
cumulant for the chiral condensate as a function of mass at $\mu_I=\mu=0$ from
our $12^3 \times 4$ simulations. Taking the point where the straight-line fit
passes through the Ising value as our estimate for the critical mass yields
$m_c=0.0264(3)$. Each of the points in this graph were obtained by averaging
the Binder cumulants taken from several $\beta$ values close to the
transition, and extrapolated to $\beta_c$ which minimizes these cumulants,
using Ferrenberg-Swendsen rewieghting \cite{Ferrenberg:1988yz}. 

\begin{figure}[htb]
\epsfxsize=4in
\centerline{\epsffile{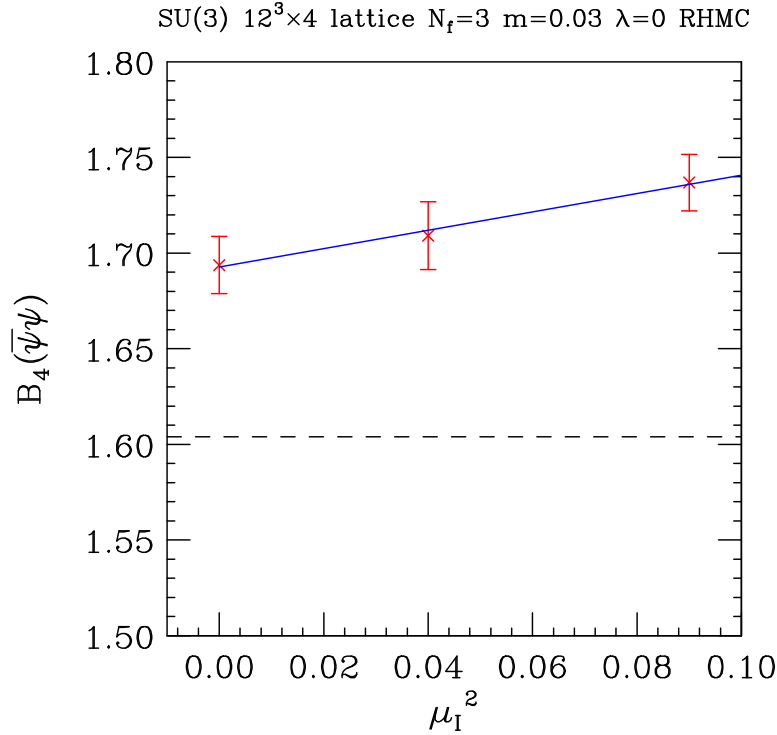}}
\caption{Binder cumulant at $T=T_c$ as a function of $\mu_I^2$ at $m=0.03$.}
\label{fig:b4m0.03}
\end{figure}
The $\mu_I$ dependence of this Binder cumulant at $\beta_c(\mu_I)$ 
is shown in figure~\ref{fig:b4m0.03}, for $m=0.03$, a little above $m_c$.
It is clear that, rather than decreasing with increasing $\mu_I$, it actually
increases slowly. Since $\beta_c$ and hence $T_c$ decrease with increasing
$\mu_I$, in physical units $m$ is actually decreasing with increasing $\mu_I$
meaning that at fixed physical $m$ the rise would be even more pronounced. The
behaviour at $m=0.035$ is very similar.

\begin{figure}[htb]
\epsfxsize=4in
\centerline{\epsffile{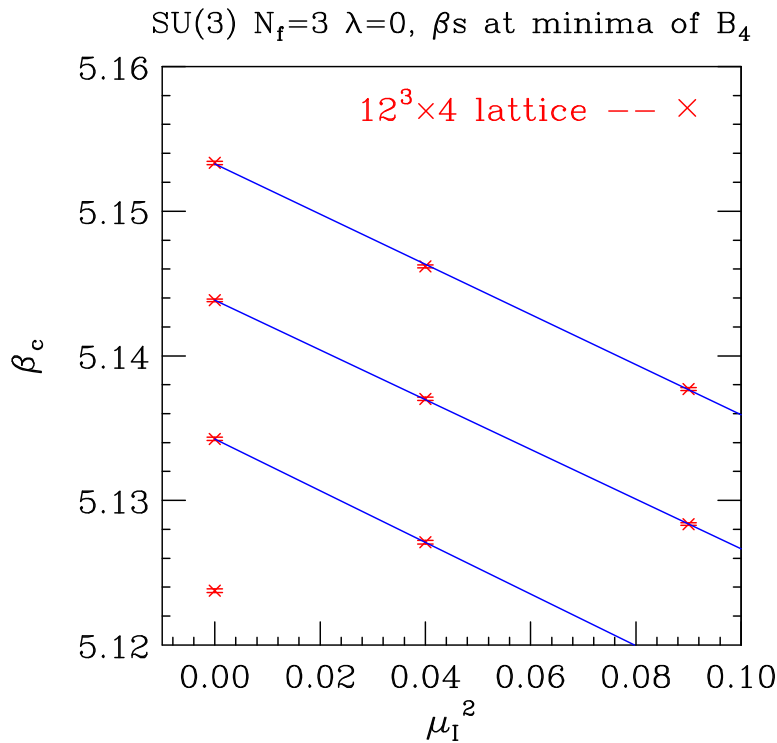}}
\caption{$\beta_c$ as functions of $\mu_I^2$. From top to bottom, $m=0.035$, 
$m=0.03$, $m=0.025$, $m=0.02$.}
\label{fig:beta_c}
\end{figure}
Figure~\ref{fig:beta_c} shows the dependence of the transition $\beta$, 
$\beta_c$, on $\mu_I$. As mentioned above, $\beta_c$ and hence the transition
temperature $T_c$ fall (slowly) with increasing $\mu_I$ as expected. The fits
shown to this preliminary `data' are:
\begin{eqnarray}
\beta_c &=& 5.15326(10)    - 0.173(2) \mu_I^2 \hspace{0.5in} m=0.035\nonumber \\
\beta_c &=& 5.14386(\:\,8) - 0.172(1) \mu_I^2 \hspace{0.5in} m=0.030\nonumber \\
\beta_c &=& 5.13426(12)    - 0.179(4) \mu_I^2 \hspace{0.5in} m=0.025\nonumber \\
\end{eqnarray}
which is in reasonable agreement with the results of de Forcrand and Philipsen
for the $\mu$ dependence of the transition temperature, obtained from analytic
continuation from imaginary $\mu$ if we make the identification $\mu_I=2\mu$.

\begin{figure}[htb]
\epsfxsize=4in
\centerline{\epsffile{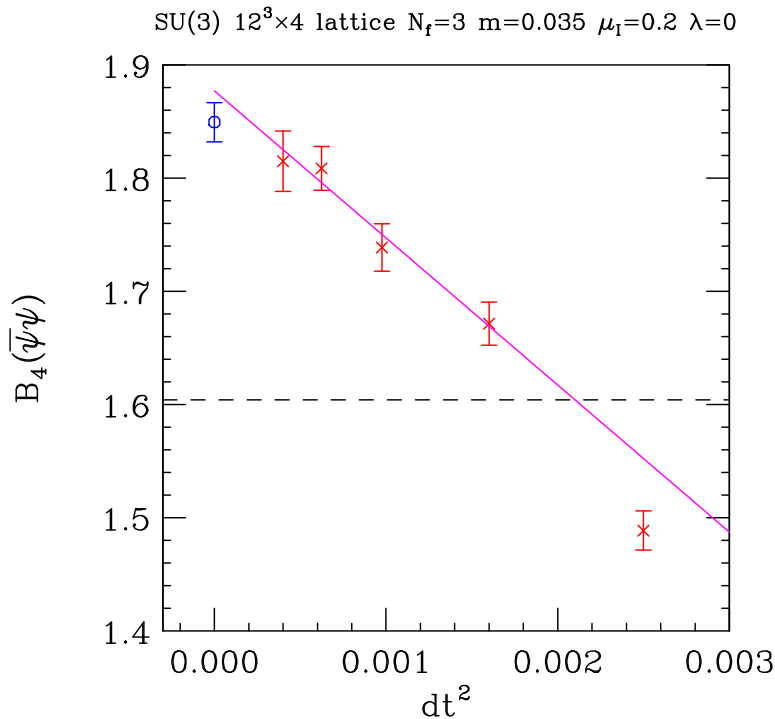}}
\caption{Binder cumulants for $m=0.035$, $\mu_I=0.2$ for HMD(R) simulations as
a function of $dt^2$ compared with that from RHMC simulations.}
\label{fig:rhmc&hmdr}
\end{figure}
Figure~\ref{fig:rhmc&hmdr} shows the $dt$ dependence of the Binder cumulants 
at the transition for $m=0.035$, $\mu_I=0.2$ in the HMD(R) simulations. The
exact RHMC result, which has no $dt$ dependence, is plotted on this graph at
$dt=0$. It is clear that the RHMC result is consistent with the 
$dt \rightarrow 0$ limit of the HMD(R) results. The actual value of $dt$ used
in the RHMC simulations was $dt=0.05$, the value of $dt$ for the rightmost
point on this graph, showing one advantage of using this new algorithm. 

\section{Conclusions}

We simulate lattice QCD with 3 flavours of staggered quarks with a small
chemical potential $\mu_I < m_\pi$ for isospin, in the neighbourhood of the
finite temperature transition from hadronic matter to a quark-gluon plasma.
Fourth order Binder cumulants are used to probe the nature of this transition
and search for the critical endpoint for masses slightly above the critical
mass for zero chemical potentials. Earlier simulations using the HMD(R)
algorithm were plagued by large finite $dt$ errors \cite{Kogut:2005yu}. We now
use the RHMC algorithm which is exact in the sense of having no finite $dt$
errors.

We measure the critical mass to be $m_c=0.0264(3)$ for $N_t=4$, in agreement
with the recent results of de Forcrand and Philipsen \cite{deForcrand:2006pv},
but considerably below earlier measurements which found values close to
$m_c=0.033$ \cite{Karsch:2001nf,Christ:2003jk,deForcrand:2003hx}. These higher
values were due to using the HMD(R) algorithm with $dt$ large enough to
produce large systematic errors.

For masses greater than $m_c$ we found that the Binder cumulant for the chiral
condensate increases with increasing $\mu_I$ and thus shows no evidence for a
critical endpoint, contrary to earlier expectations. This also agrees with the
observations of de Forcrand and Philipsen \cite{deForcrand:2006pv} for finite
$\mu$, emphasizing the similarities between finite $\mu$ and finite $\mu_I$
for small $\mu$,$\mu_I$, near the finite temperature transition. On these
relatively small lattices ($12^3 \times 4$), we really should minimize the
Binder cumulant of linear combinations of the chiral condensate, the plaquette
and the isospin density to obtain the desired eigenfield of the
renormalization group equations, to draw reliable conclusions
\cite{Karsch:2001nf} \footnote{We thank Frithjof Karsch for reminding us of
this fact}.

We end with the observation that we have used RHMC simulations where we do not
know a positive lower bound for the spectrum of the quadratic Dirac operator.
This is done by choosing a speculative lower bound and justifying our choice
a postiori. We refer the reader to our recent paper on this subject
\cite{Kogut:2006jg}.

\section*{Acknowledgements}

JBK is supported in part by a National Science Foundation grant NSF
PHY03-04252. DKS is supported by the U.S. Department of Energy, Division of
High Energy Physics, Contract W-31-109-ENG-38. We thank Philippe de Forcrand
and Owe Philipsen for helpful discussions. Simulations are performed on Jazz
at Argonne, Tungsten and Cobalt at NCSA, Bassi and Jacquard at NERSC,
Data\-Star at SDSC and PC's in Argonne HEP.

\end{document}